\documentclass[%
 reprint,
 amsmath,amssymb,
 aps,
]{revtex4-2}

\usepackage{graphicx}
\usepackage{dcolumn}
\usepackage{bm}
\usepackage{caption}
\usepackage{subcaption}
\usepackage{dsfont}
\usepackage[section]{placeins}

\begin{document}

\preprint{APS/123-QED}

\author{Anurag Singh}
\author{Merle I. S. Röhr}
\email{merle.roehr@uni-wuerzburg.de}
\affiliation{Institute of Physical and Theoretical Chemistry, Julius-Maximilians-Universität Würzburg, Emil-Fischer-Sr. 42, 97074,Würzburg,Germany}
\affiliation{Center for Nanosystems Chemistry, Julius-Maximilians-Universität Würzburg,Theodor-Boveri Weg,97074,Würzburg,Germany}

\title[An \textsf{achemso} demo]
  {Formation Pathways of the Spin-Correlated, Spatially Separated $^{1}$(T...T) State in the Singlet Fission Process of Perylene Diimide Stacks}


\begin{abstract} 
For the theoretical screening of Singlet Fission (SF) rates in molecular aggregates, commonly dimer model systems are employed. However, there is experimental evidence, that the SF process proceeds from the $^{1}$(TT) state via an triplet-triplet energy transfer process to a further intermediate: a $^{1}$(T...T) state with two non-adjacent, spin-correlated triplets, which cannot be captured by the dimer models. In this work, we extend Michl's diabatic frontier orbital model to trimer systems, for which we automatically generate the diabatic two and three-center  couplings  using symbolic algebra. We apply this method to study the packing dependence of the $^{1}$(T...T) formation in the perylene diimide (PDI) trimer stack. We find that efficient triplet-triplet energy transfer is facilitated by structural motifs for which also significant excimer character can be observed. Furthermore, the coupling shows a local maximum for the structural motif that has been assigned to efficient  $^{1}$(TT) population.  Employing second order perturbation theory, we study the interference of the individual electronic pathways that arise in the PDI trimer system, allowing us to reproduce the packing dependence of the SF rates derived from the Redfield simulations in the work by Mirjani et. al. (Phys. Chem. C 2014, 118, 26, 14192–14199). 
\end{abstract}

\maketitle


\section{Introduction}\label{sec1}

Singlet fission (SF) is a process occurring in molecular systems, in which a singlet exciton is converted into two triplets.\cite{smith2010singlet} As SF offers the opportunity to overcome the thermalization loss in solar energy technologies, the design of efficient SF materials is an active field of research that crucially builds on the understanding how multiexcitons can be generated and controlled.\cite{lee_singlet_2013,xia_singlet_2017} It is commonly accepted that the SF process proceeds from an initially excited singlet exciton state via the transient formation of a correlated triplet exciton pair $^{1}(\text{TT})$ intermediate with an overall singlet spin multiplicity, either via a direct or a charge transfer mediated pathway, however, the fate of the spin-correlated triplets is far less explored. \cite{Miyata2019,sanders_understanding_2019,hudson_what_2022} The populated $^{1}(\text{TT})$ state has been commonly assumed to disentangle by spin decoherence effects, forming two independent triplets  T +  T.\cite{Casanova2018} However, Pensack et. all. reported on a femtosecond transient absorption study of the SF process in pentacene, providing evidence that the intermediate $^{1}(\text{TT})$ state further evolves to a second intermediate, forming a specially separated, spin correlated triplet exciton pair $^{1}(\text{T...T})$.\cite{Pensack2016} Based on their findings, they established a three-step kinetic scheme of the SF process:

\begin{equation}
    S_{0} + S_{1} \rightarrow [ ^{1}(\text{TT}) \rightleftharpoons \text{   } ^{1}(\text{T}...\text{T})] \rightleftharpoons  T + T
\end{equation}

The $^{1}(\text{T...T})$ state can be considered as an electronically uncorrelated, but yet spin entangled triplet pair on non-adjacent molecules, that can be formed from a $^{1}(\text{TT})$ state upon a triplet–triplet energy (Dexter) transfer mechanism. \cite{Scholes2015, Lee2018} For deeper information, we refer the reader to the review by Zhu and coworkers\cite{Miyata2019}.
We wish to emphasize that the introduced three-step mechanism offers a so far unexplored strategy in the design of an efficient SF process: that is, the harvesting of the $^{1}$(T...T)state in order to steer the kinetics of the process. Interestingly, He et. al. have recently reported on the direct harvesting of a bound triplet pair.\cite{he_direct_2022} 
Furthermore, in the computational screening of new SF materials, the transfer integral for the $^{1}$(TT) to $^{1}$(T...T) transition should be taken into account as well as its packing dependence.
However, this aspect is underexplored from theoretical side. Abraham et. al. have delivered an expression of the respective triplet–triplet energy transfer integral derived from the Spin Hamiltonian in the Heisenberg picture and studied its dependence on the packing motif in tetracene employing ab initio calculations. \cite{abraham_revealing_2021} Taffet et. al. carried out highly accurate single-point computations of the noninteracting $^{5}$TT state in the dimer that can be considered as a proxy for the spatial product of the $^{1}$(T...T) state, studying tetracenes as well as carotenoids. \cite{taffet_overlap-driven_2020} 

The existence of the described intermediates has already been proposed in the 70s by Frankevich et. al. based on the Reaction Yield Detected Magnetic Resonance Spectra they recorded for the tetracene crystal, \cite{Frankevich1978} while Chan et. al. has fomulated a similar kinetic scheme upon the observation of two distinct multiexciton states in a pentacene/fullerene bilayer system using femtosecond nonlinear spectroscopies.\cite{Chan2011} 
The formation of an $^{1}$(T...T) state in pentacene was also studied by others \cite{Grieco2017,SrimathKandada2014} and has been further observed in crystalline rubrene \cite{Baronas}, polycrystalline hexacene \cite{Qian} as well as in linear oligomers and linked dimers of perylene\cite{Korovina2020,korovina_singlet_2016}, each representing well-established SF molecules.  Another prominent candidate for studying SF effects is perylene diimide (PDI), that offers the appealing possibility to effectively control its packing arrangement in thin films introducing different functional groups at their imide positions,\cite{le2018singlet,Eaton2013} while the introduction of linkers on these positions allows for the production of well-defined model compounds of stacked dimer and trimer oligomers that have provided valuable insights into the packing structure dependence of SF efficiency and underlying mechanisms.\cite{Chen2018,Hong2020, Hong2022}  Studies suggest a CT-mediated mechanism promoted by intermolecular orbital effects, while the role of a (transient) excimer formation is under debate.\cite{mirjani2014theoretical,miller2017modeling,Hong2020}  A recent combined theoretical and experimental study on a linked trimer stack of PDIs revealed two SF channels that depend on the chosen wavelength of the initial excitation: a fast pathway that has been assigned to proceed via a virtual CT coupling, initiated by the excitation of a red band, and a (slower) pathway that is assumed to proceed via transient formation of the excimer, initialized upon the excitation of the main absorption band.\cite{Hong2022} However, to the best of our knowledge, no experimental ( and theoretical work) has targeted the mutual role of the specially separated $^{1}$(T...T) state in the SF process of PDIs. In this work, we deliver a theoretical study on the formation pathways and their packing dependence of the specially separated, spin-correlated triplet pair $^{1}$(T...T) state in the PDI trimer system, with the aim to stimulate further both experimental and theoretical studies in this field to fully explore the role of the $^{1}$(T...T) state in the SF process.

In the past, theoretical studies of the SF process and its packing dependence have often focused on molecular dimers. \cite{Zimmerman2010,Teichen2012, Greyson2010,Berkelbach2013,Feng2013,Casanova2014,Zeng2014, renaud2013mapping,Zaykov2019} In this regard, the diabatic frontier orbital (FO) model as introduced by Michl \cite{smith2010singlet}, in which the local excitation (LE), Charge Transfer (CT) as well as the singlet-correlated triplet pair excitation in the basis of the monomer HOMOs and LUMOs  are considered is of frequent use. Approximating the couplings between these quasi-diabatic states by expressions of the Fock matrix, the model has been proven particularly suitable for the fast screening of the packing dependence of the SF process in molecular dimer systems. \cite{Ryerson2019,Zaykov2019}   Berkelbach et. al. delivered an ab initio parametrized implementation of the FO model and carried out periodic calculations in order to study the SF process in the crystalline phase, accounting for couplings between next neighbored molecules,\cite{berkelbach_microscopic_2013,berkelbach_microscopic_2013-1,berkelbach_microscopic_2014}, Li et al. delivered an ab-initio parametrized, extended exciton model for SF that considers the FMO model and couplings between adjacent molecules, studying pentacene clusters, while Nakano et. al. constructed linear aggregates with defined extended size based on the dimer parametrization.\cite{nakano_quantum_2019}  Miryani et.al. screened the packing dependence of the SF rate by Redfield simulations, in which the FO dimer model is coupled to a bath, in order to account for the crystal environment. However, all these implementations are inherently insufficient for the investigation of the formation of the $^{1}$(T...T) state as for such studies,  three-center couplings have to be considered. For such studies, an explicit trimer system, in which couplings between all monomers are tsken into account, has to be employed as a minimal model. 

In this work, we generalize the FO model for the description of molecular trimer systems, accounting for couplings between all monomers and employ second order perturbation theory to derive the Transition Probability of the formation of the correlated triplet pair state, which we scan along a vertical and a longitudinal slipping mode and study the electronic pathways of the $^{1}$(T ... T) formation. We further determine the relative importance of the electronic pathways that arise in the trimer system and study the packing structure dependence of the individual formation pathways, identifying optimal stacking geometries for the SF process. In our algorithm, matrix elements are automatically generated using symbolic algebra, allowing an efficient calculation of products of matrix elements.

\section{Method}
\begin{figure}
    \centering
    \includegraphics[width=1\linewidth]{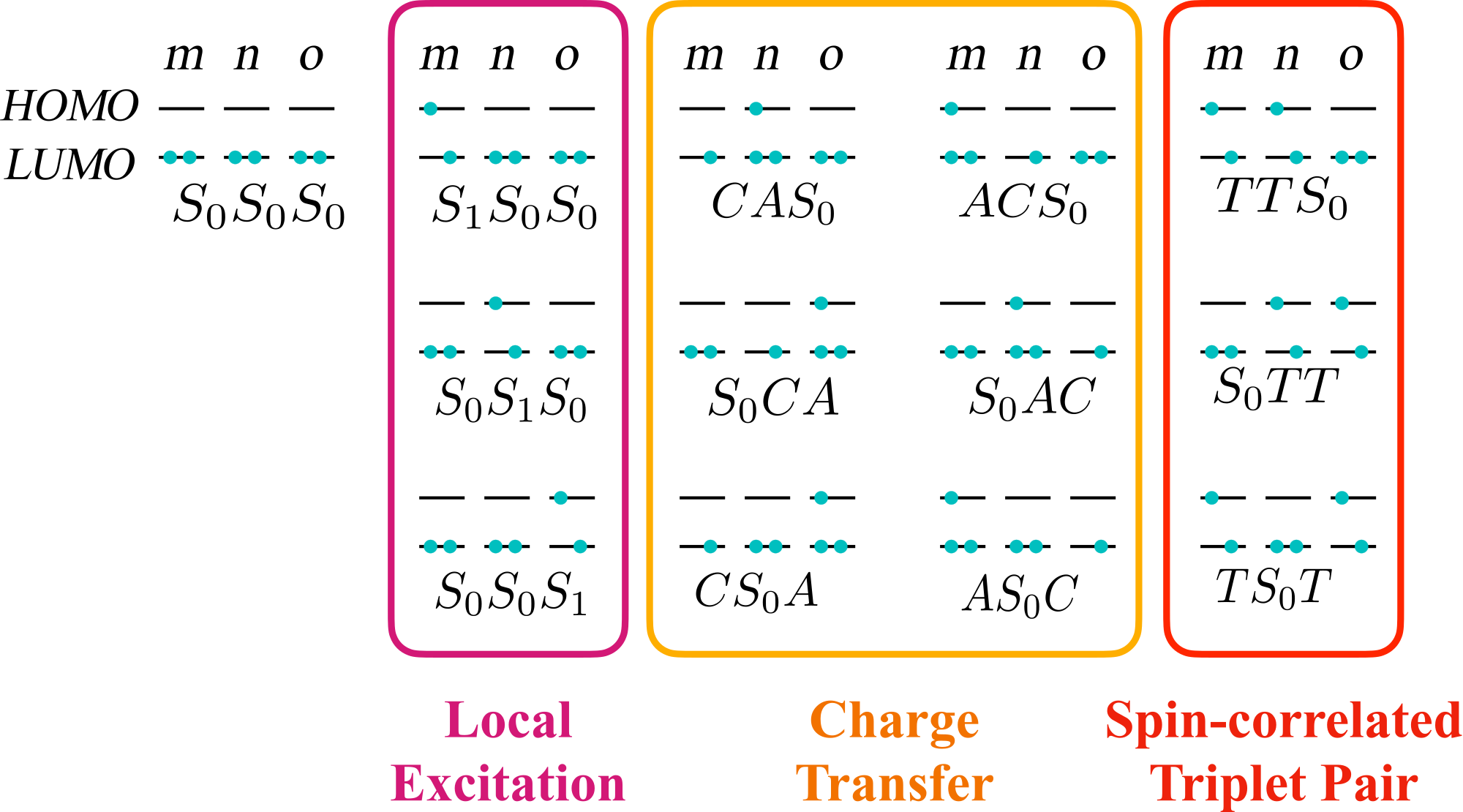}
    \caption{Diabatic states considered in the trimer model Hamiltonian.}
    \label{fig:FrontierModel}
\end{figure}

In the following, the details of the model are presented: We construct a stack consisting of three monomers $m$, $n$ and $o$. Considering the highest occupied molecular orbitals (HOMO) and lowest unoccupied molecular orbitals (LUMO) on each monomer, the model describes a system with 6 electrons in the 6 orbitals $h_{m}$, $l_{m}$, $h_{n}$, $l_{n}$, $h_{o}$ and $l_{o}$, where $h$ and  $l$ represent HOMO and LUMO, respectively, and their subscripts represent the monomer they belong to. 
The orbitals are considered to be real, normalized and orthogonal to each other. 
Each monomer I=$m,n,o$ has 5 states, the ground state $S_{0}^{I}$, the singlet excited state $S_{1}^{I}$ , a cationic state $C^{I}$, an anionic state $A^{I}$, and a triplet state $T^{I}$. Assuming an overall singlet multiplicity of the system,  we construct 13 diabatic states of the trimer, giving rise to one ground state (GS), three locally excited singlet states (LE), six charge transfer states (CT) and three correlated paired triplet states ($^{1}\text{TT}$) with overall singlet multiplicity, as illustrated in figure \ref{fig:FrontierModel}. Notice, that the doubly excited state ($^{1}\text{S}_{1}\text{S}_{1}$) is excluded from further consideration, as the energy of this configuration is expected to be significantly higher.  It is emphasised, that in contrast to a dimer system, in the trimer system configurations arise, in which the cation-anion pair as well as  the  spin-correlated triplets, are spatially separated by an "innocent" monomer. 
 The resulting diabatic Hamiltonian of the model-CI system is then generated, for which the expressions for the matrix elements are derived employing our program given in ref \cite{githubGitHubRoehrlabSymbolicCI}. 
In the implementation, the second quantization formalism is applied, leading to the following expression for the Hamiltonian:
 
\begin{equation}
\begin{split}
H = \sum_{i,j = 1}^{d}\sum_{s_{i},s_{j} = \alpha,\beta}T_{ij}c_{i,s_{i}}^{\dagger}c_{j,s_{j}} \\ +  \frac{1}{2}\sum_{i,j,k,l = 1}^{d}\sum_{s_{i},s_{j},s_{k},s_{l} = \alpha,\beta}V_{ijkl}c_{i,s_{i}}^{\dagger} c_{j,s_{j}}^{\dagger}c_{k,s_{k}} c_{l,s_{l}}
\end{split}
\label{eq:hamiltonian}
\end{equation}
where ${T}$ and ${V}$ represent the one electron and 2-electron integrals respectively and $c^{\dagger}$ and $c$ are the creation and annihilation operators.

In that way the matrix elements of the Hamiltonian can be generally expressed as
\begin{equation}
\begin{split}
    \left\langle\Psi_{m}\right \vert H\left \vert \Psi_{n}\right\rangle = \sum_{i,j = 1}^{d}\sum_{s_{i},s_{j} = \alpha,\beta}T_{ij}\left\langle\Psi_{m}\right \vert c_{i,s_{i}}^{\dagger}c_{j,s_{j}}\left  \vert \Psi_{n}\right\rangle \\ + \frac{1}{2}\sum_{i,j,k,l = 1}^{d}\sum_{\substack{s_{i},s_{j},s_{k},s_{l} \\= \alpha,\beta}}V_{ijkl}\left\langle\Psi_{m}\right \vert c_{i,s_{i}}^{\dagger} c_{j,s_{j}}^{\dagger}c_{k,s_{l}} c_{l,s_{k}}\left \vert \Psi_{n}\right\rangle
    \label{eq:coup}
\end{split}
\end{equation}
A crucial aspect of our algorithm is the choice of the Jordan-Wigner representation \cite{Jordan1928} of the fermionic creation and annihilation operators in order to automatically generate the model CI-Hamiltonian matrix and provide analytic expressions for the latter. The second-quantized operators are constructed by symbolic Kronecker products of the Pauli spin matrices corresponding to the spin orbitals, and the matrix elements are generated using the symbolic algebra programs. The Jordan-Wigner representation of the fermionic operators is given by:
\begin{equation}
    c^{\dagger} =  \sigma_{z} \otimes ... \otimes \sigma_{z} \otimes \sigma^{-}\otimes \mathds{1} \otimes ... \otimes \mathds{1}
\end{equation}
\begin{equation}
    c =  \sigma_{z} \otimes ... \otimes \sigma_{z} \otimes \sigma^{+}\otimes \mathds{1} \otimes ... \otimes \mathds{1}
\end{equation}

where  $\sigma_{z}$ is the Pauli spin matrix corresponding to the z-component of the spin and $\sigma^{\pm}=\sigma_{x}\pm \iota \sigma_{y}$. Using these basic operators, spin adapted combinations of the fermionic operators can be constructed.

Using the operators formed as matrix of symbols, the one- and two-electron operators are obtained by performing tensor products of the basic operators. 
In this way, analytic expressions for the matrix elements involving configurations with arbitrary excitation level can be symbolically generated and utilized both for theoretical analysis of the couplings as well as for their numerical evaluation. 

\begin{figure}[h]
    \centering
    \includegraphics[width=1\linewidth]{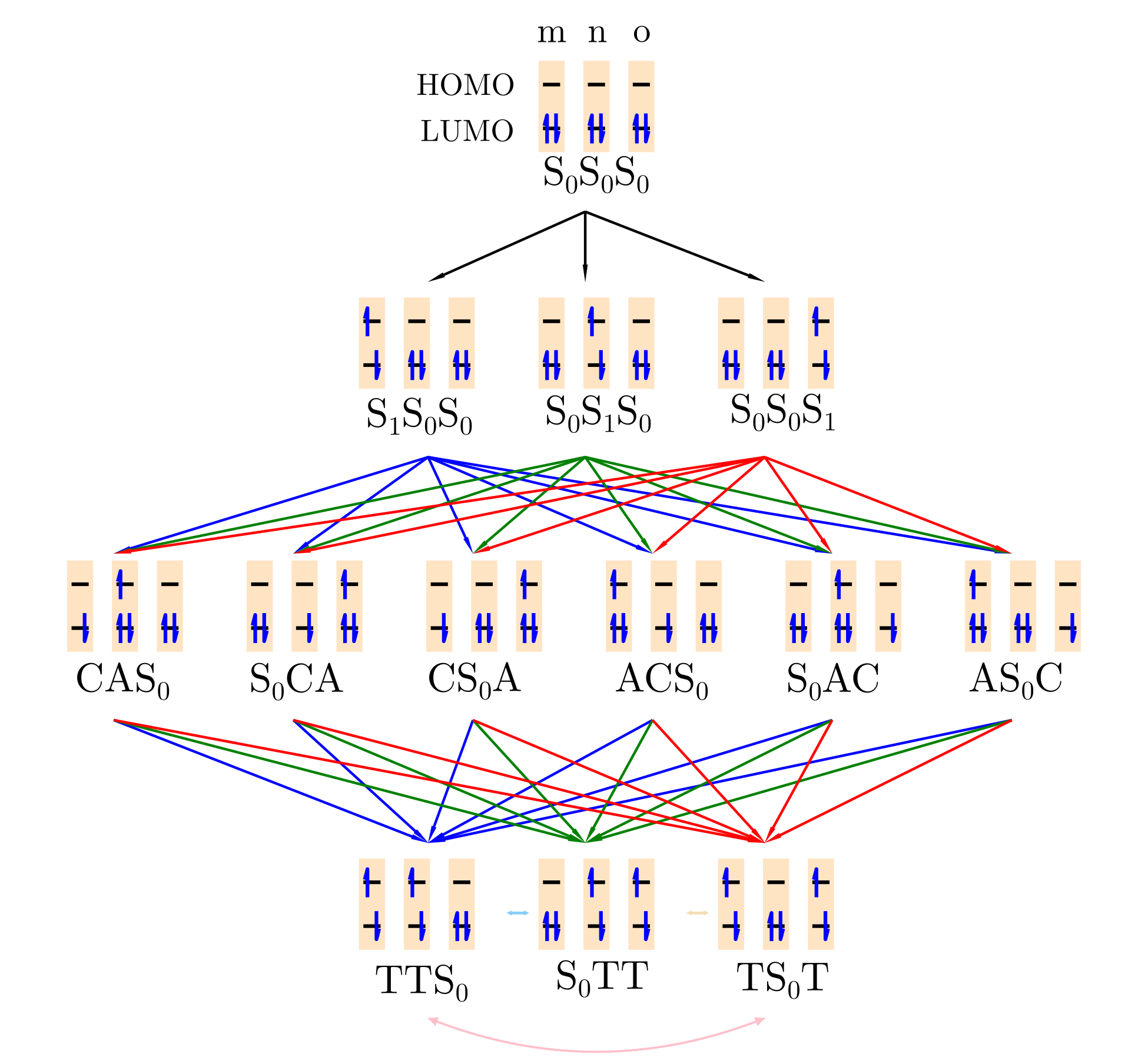}
    \caption{ b) CT mediated pathways of $^{1}$(TT) and $^{1}$(T...T) states formation}
    \label{fig:pathways}
\end{figure}

\begin{figure*}
\centering
     \begin{subfigure}{1.0\textwidth}
         \centering
         \includegraphics[width=\textwidth]{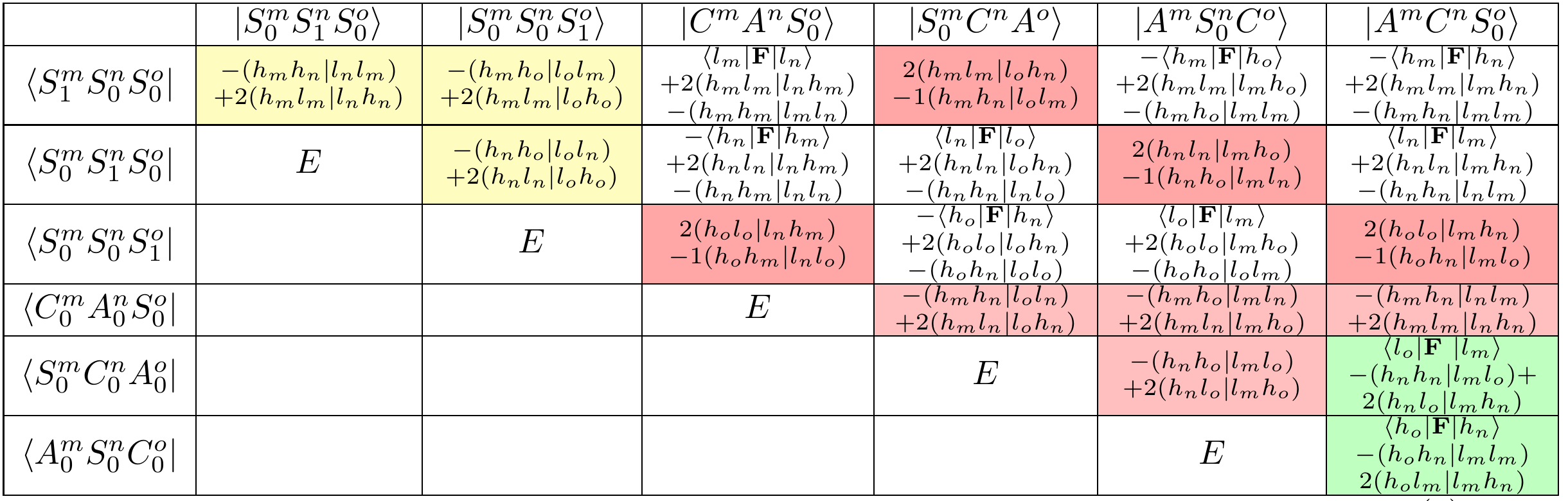}
         \label{fig:y equals x}
     \end{subfigure}
     \hfill
     \begin{subfigure}{1.0\textwidth}
         \centering
         \includegraphics[width=\textwidth]{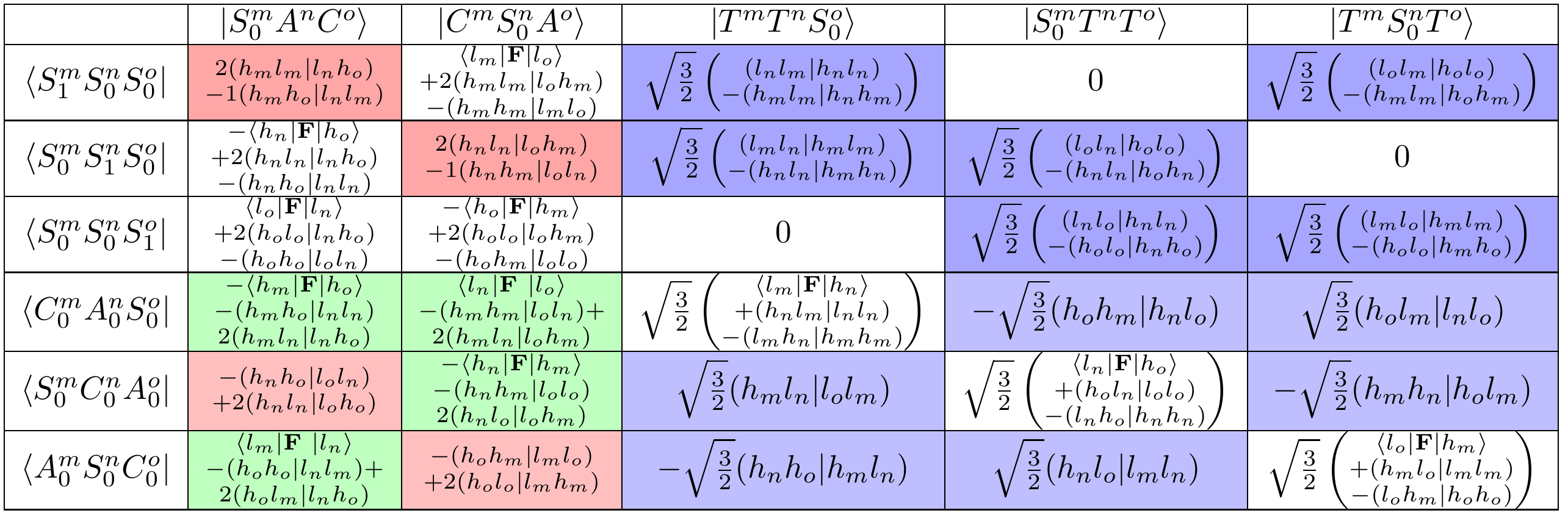}
         \label{fig:three sin x}
     \end{subfigure}
     \hfill
     \begin{subfigure}{1.0\textwidth}
         \centering
         \includegraphics[width=\textwidth]{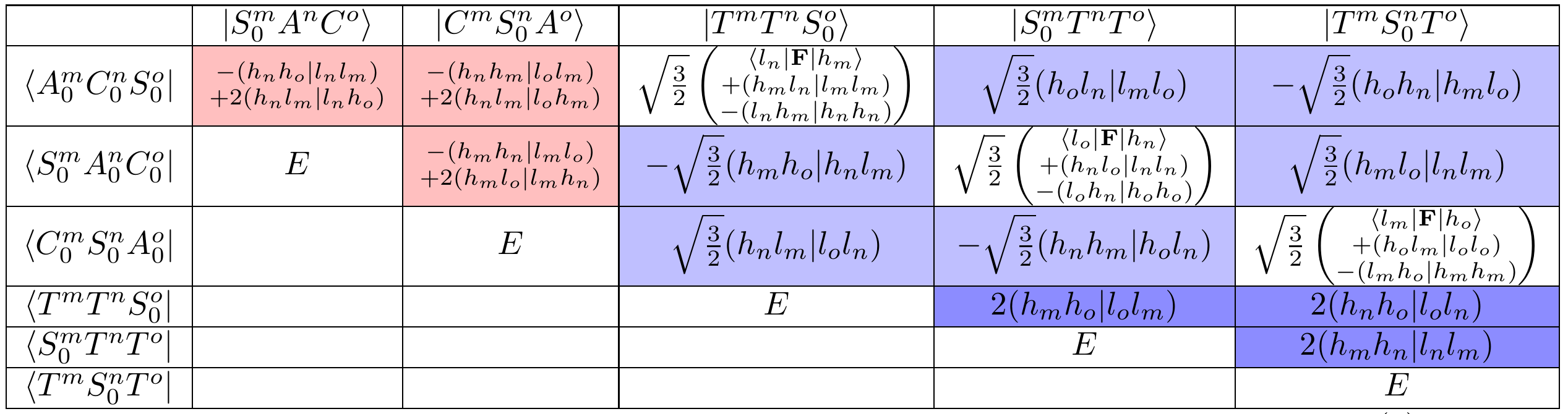}
         \label{fig:five over x}
     \end{subfigure}

    \caption{Diabatic Hamiltonian of the trimer system, considering the HOMO and LUMO of monomers. The diabatic states involve 3 Singlet exciton states 6 Charge transfer states and 3 correlated triplet states as illustrated in Fig. 1. The matrix elements are represented in terms of the Fock operator and electron repulsion integrals (ERI) in molecular orbital basis.  }
    \label{fig:Hamiltonian}
\end{figure*}

In this work, we approximate the formation rate from a chosen initial diabatic state to the chosen final state by Fermi's golden rule, employing second-order perturbation theory. In that way, a transition probability can be calculated according to.  
\begin{equation} 
    P_{a \rightarrow b} = \left \vert \sum_{a \neq b} \sum_{ q \neq b } \sum_{a} \frac{ \langle a \vert W_{t} \vert q \rangle\langle q \vert W_{t} \vert b \rangle}{(E_{b} - E_{a})(E_{q} - E_{b})}  \right \vert^{2},
    \label{eq:mediatedProbability}
\end{equation}
where  $ a$  and  $b$ are the initial and final diabatic states, respectively, $q$ is the intermediate diabatic state  and $W_{t}$ is the perturbing  Hamiltonian. $E_{a}, E_{b}$ and $E_{q}$ are the energy of diabatic states $a,b$ and $q$ respectively.
Notice, that the density of states, that depends on the vibrational structure, is neglected.
For a given pathway of the SF process where the transition is occurring from any local exciton state $LE_{a}$ to any final triplet pair state $^{1}\text{(TT)}_{b}$ , involving any charge transfer state $CT_{i}$, the transfer rate depends on the transfer probability $^{i}T_{LE \rightarrow ^{1}\text{(TT)}}$ according to equation \ref{eq:mediatedProbability_CTi}. 
\begin{equation} 
    ^{i}T_{LE_{a} \rightarrow ^{1}\text{(TT)}_{b}} = \frac{ \langle LE_{a} \vert W \vert CT_{i} \rangle\langle CT_{i} \vert W \vert ^{1}\text{(TT)}_{b} \rangle}{(E_{^{1}\text{(TT)}_{b}} - E_{LE_{a}})(E_{CT_{i}} - E_{^{1}\text{(TT)}_{b}})}
    \label{eq:mediatedProbability_CTi}
\end{equation}
Adding up $^{i}T_{LE_{a} \rightarrow ^{1}\text{(TT)}_{b}}$ for all possible CT states results in a total transfer probability $T_{LE _{a} \rightarrow ^{1}\text{(TT)}_{b}}$:      
\begin{equation} 
    T_{LE _{a} \rightarrow ^{1}\text{(TT)}_{b}} = \sum_{ CT_{i} } \frac{ \langle LE_{a} \vert W \vert CT_{i} \rangle\langle CT_{i} \vert W \vert ^{1}\text{(TT)}_{b} \rangle}{(E_{^{1}\text{(TT)}_{b}} - E_{LE_{a}})(E_{CT_{i}} - E_{^{1}\text{(TT)}_{b}})}  
    \label{eq:mediatedProbability_CT}
\end{equation}
We assume, that the energies of the individual CT states do not differ, as they are derived from the ionisation potential of the monomers. Thus, $E_{CT_{i}} - E_{^{1}TT_{b}}$ are approximated to be same for all $i$. In that way, the expression reduces to
\begin{equation} 
    T_{LE _{a} \rightarrow ^{1}\text{(TT)}_{b}} \approx \sum_{ CT_{i} }\langle LE_{a} \vert W \vert CT_{i} \rangle\langle CT_{i} \vert W \vert ^{1}\text{(TT)}_{b} \rangle  \times c
    \label{eq:mediatedProbability_CTf}
\end{equation}
where
\begin{equation}
    c = \frac{1}{(E_{^{1}TT_{b}} - E_{LE_{a}})(E_{CT_{i}} - E_{^{1}TT_{b}})}
\end{equation}

\section{Computational Details}\label{ComDet}
The molecular geometry of the PBI monomer has been optimised using DFT, employing the CAM-B3LYP functional and the 6-31g basis set.
For the calculation of the Fock matrix for the trimer system in the atomic orbital basis ($\text{F}^{\text{AO}}$), the  Hartree-Fock method along with the 6-31g and def2-svp basis set has been employed, as implemented in PYSCF. The coefficient matrix of the monomers were constructed from separate Hartree-Fock calculations. A combined coefficient  matrix of the trimer has been generated with the monomer's coefficient matrices as diagonal elements. The Fock operator in molecular orbital basis $\text{F}^{\text{MO}}$, as it appears in the matrix elements given in figure \ref{fig:Hamiltonian} is calculated by transforming $\text{F}^{\text{AO}}$ according to :
\begin{equation}
     \text{F}^{MO}_{a,b}  = \langle \psi_{a} \vert \text{F} \vert \psi_{b} \rangle = \langle \text{C}_{a}^{\dagger} \vert \text{F}^{\text{AO}} \vert \text{C}_{b} \rangle = \sum_{j}\sum_{i}C^{a}_{i} \text{F}_{i,j}^{\text{AO}} C^{b}_{j} 
    \label{eq:fockIntegral}
\end{equation}
where $\text{C}_{a}$ is the coefficient vector of molecular orbital $a$ . The electron repulsion integrals (ERI) in the molecular orbital basis are calculated by transforming their counterparts in the AO basis as 
\begin{equation}
    \langle \psi_{a} \psi_{b} \vert \psi_{c} \psi_{d} \rangle = \sum_{i}\sum_{j}\sum_{k}\sum_{j}C^{a}_{i}C^{b}_{j}C^{c}_{k}C^{d}_{l}\langle i j \vert kl\rangle
    \label{eq:twoI}
\end{equation}
The matrix elements of the dibatic Hamiltonian as given in Fig. \ref{fig:Hamiltonian} are derived using our code that uses SymPy, a Python library for symbolic mathematics \cite{10.7717/peerj-cs.103}.

To scan the diabatic coupling elements in the trimer system, we constructed a trimer stack of the three monomers, where consecutive monomers were stacked at an interplanar distance of $3.4 \text{\AA}$ . The first monomer was kept stationary, and the second and third monomers were allowed to slip along both the longitudinal and transversal axes from $0 \text{\AA}$ to $4.0 \text{\AA}$.

We performed the scanning in steps of $0.1 \text{\AA}$, where the slipping of the third monomer was twice that of the second monomer in each step. The slipping was performed in such a way that the consecutive monomers were slipped along the same vector. This ensured that each monomer felt the same interaction with its neighboring monomer.

By calculating the diabatic Hamiltonian in each step, we systematically explore the diabatic coupling elements as a function of the slipping mode.

\begin{figure}
    \centering
    \includegraphics[width=1.0\linewidth]{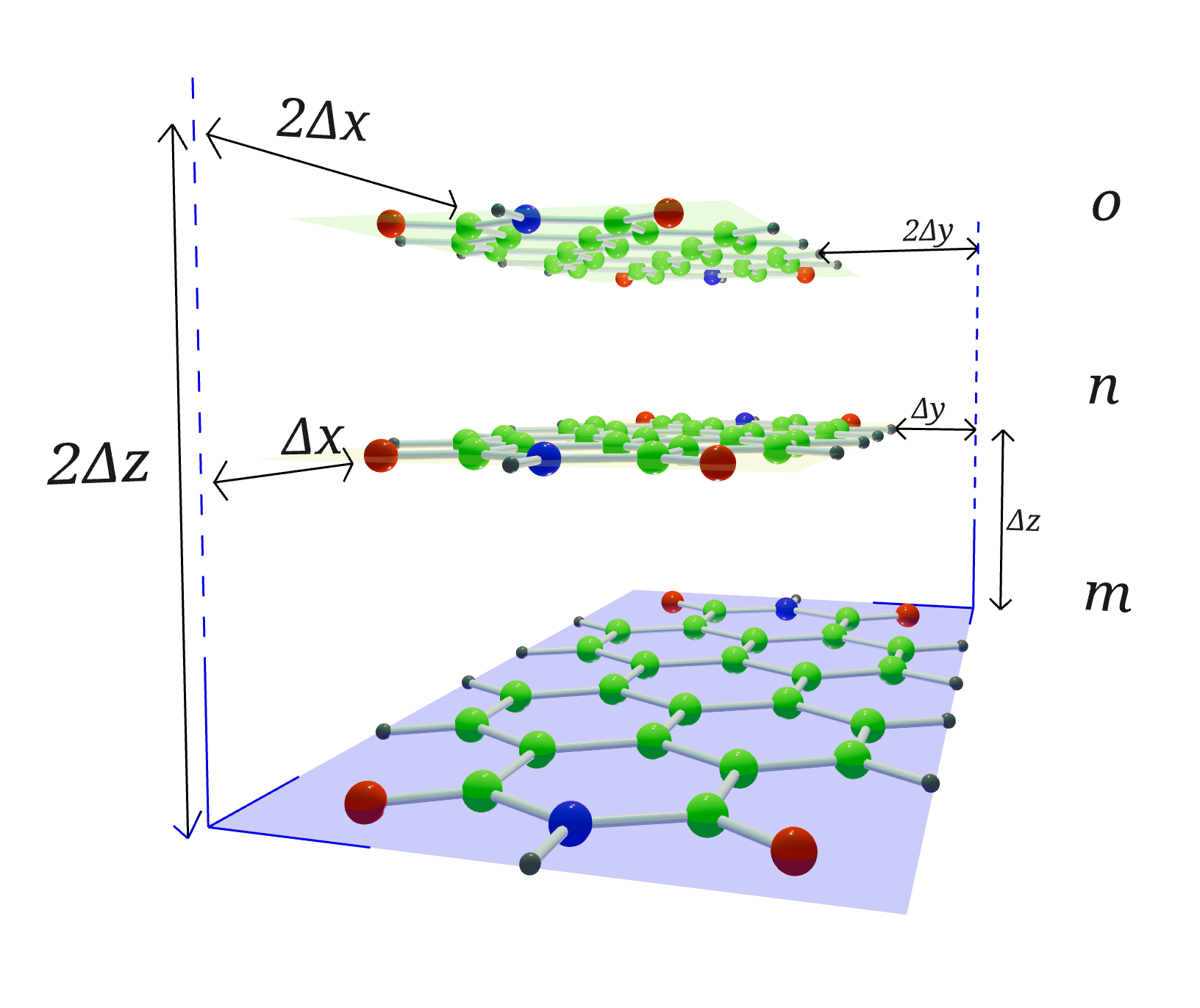}
    \caption{ PDI trimer model}
      \label{fig:CancellingMatrix}
\end{figure}

\section{Results}\label{Result}

\begin{figure*}
    \centering
    \includegraphics[width=1.0\linewidth]{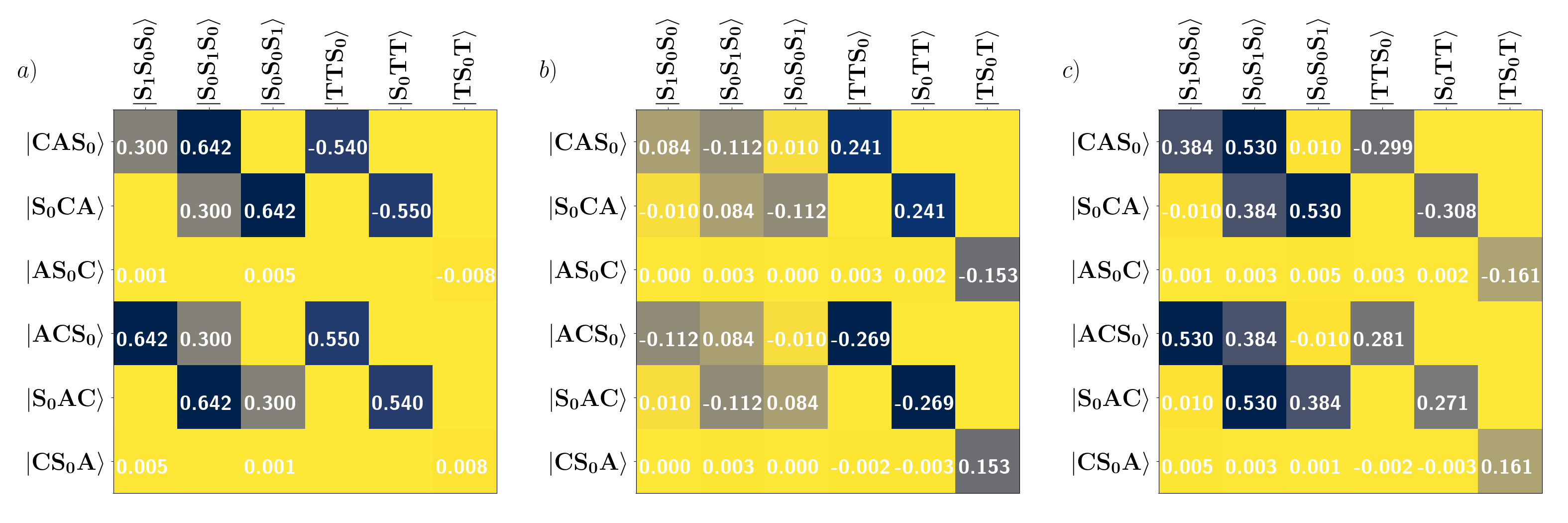}
    \caption{Matrix elements at X=2.9 $\text{\AA}$, Y= 0.0 $\text{\AA}$, a) only Fock matrix values, b) Only the ERI values, c)Fock + ERI }
    \label{fig:PBISlippingStructure}
\end{figure*}
\begin{figure}[h]
    \centering
    \includegraphics[width=1.0\linewidth]{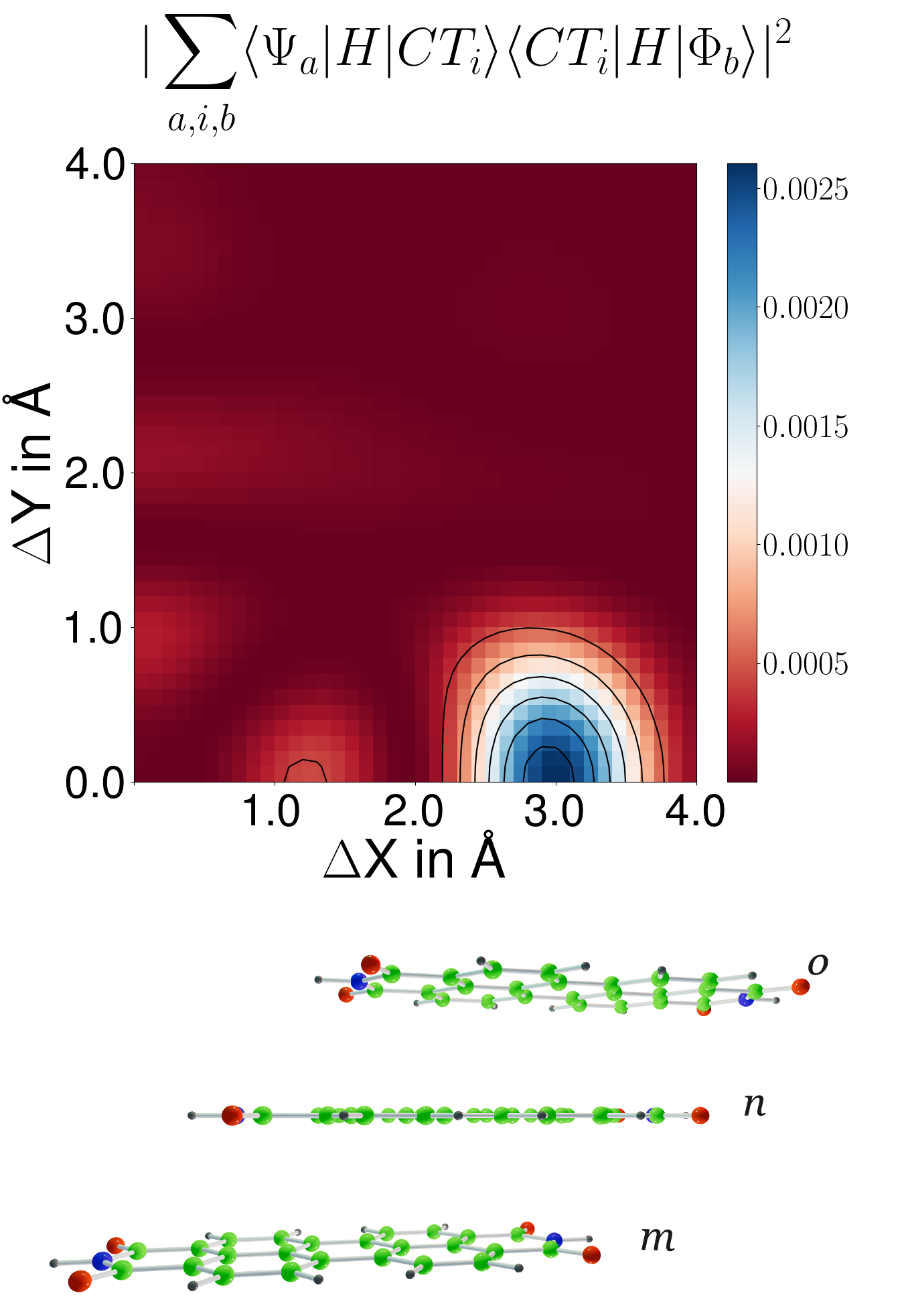}
    \caption{a) Scan of the Transition Probability for the $^{1}(\text{TT})$ formation (Sum over all transfer probabilities from LT ($\Psi$) to $^{1}(\text{TT})$ ($\Phi$)) .  b) Packing motif of the trimer system at $\Delta x = 2.9 \text{\AA}$ and  $\Delta y = 0.0 \text{\AA}$ }.
    \label{fig:ScanGeometry}
\end{figure}
\begin{figure*}
    \centering
    \includegraphics[width=1.0\linewidth]{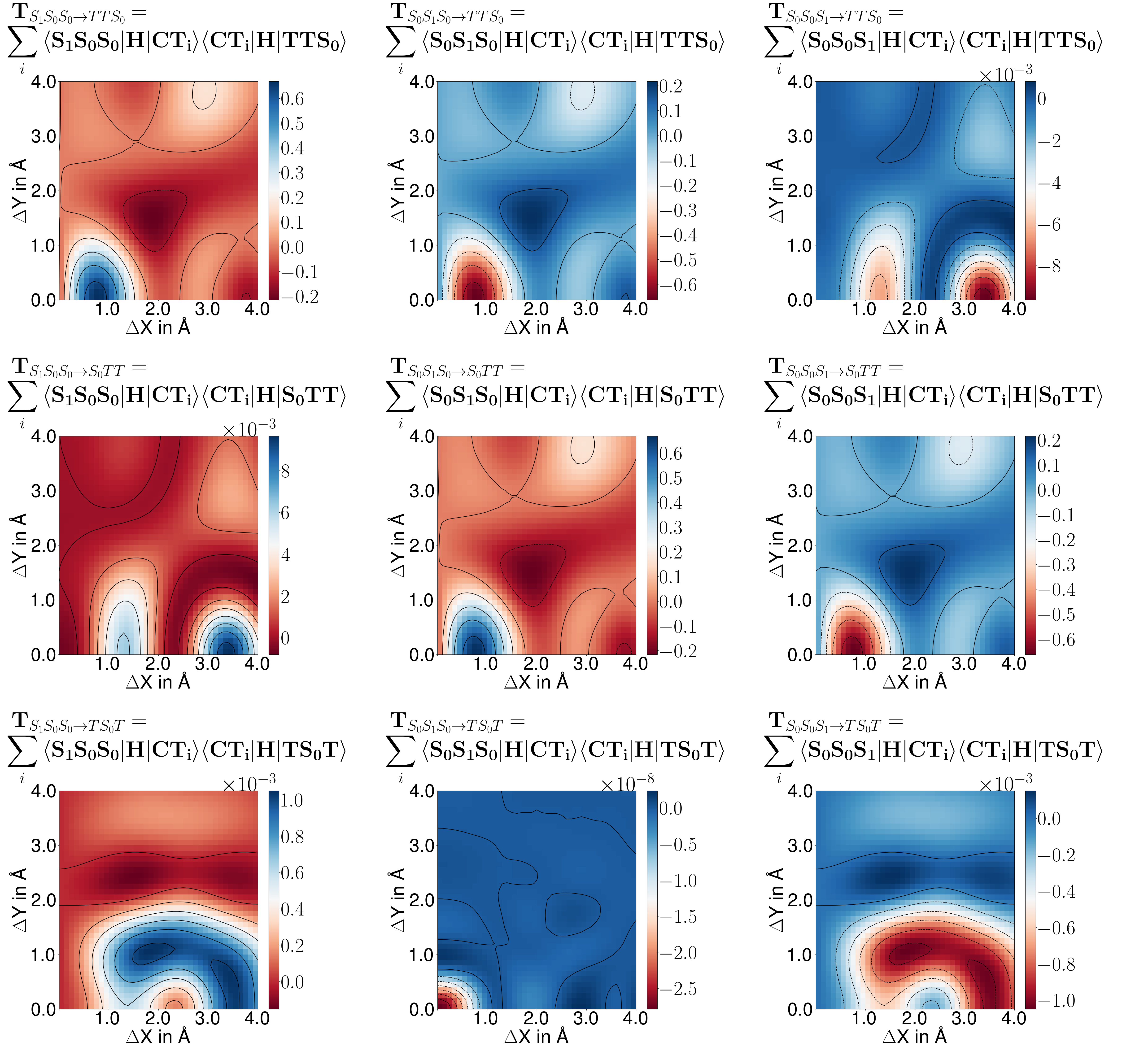}
    \caption{Plots of transfer probability from individual LE to individual $^{1}(\text{TT})$ or $^{1}(\text{T...T})$ state, summed  over all intermediate CT states. Each column has common initial LE state and each row has common final TT state. }
    \label{fig:plotMediated}
\end{figure*}

We generated the diabatic Hamiltonian for the trimer system, based on the procedure delineated in the "Methods" section. Notice, that our approach differs from Michl's Simple Model (SM) \cite{Buchanan2017}, since in the Simple Model, the coupling of the locally excited (LE) states with the $^{1}$(TT) states is neglected, while our method specifically constructs these matrix elements. While in the SM, the matrix elemts are essentially broken down to 1 electron integrals, we employ the full analytic expression derived for the indiviual matrix elements.  Furthermore, in the trimer Hamiltonian we consider the configurations $^{1}\vert TS_{0}T \rangle$ as well as the  $\vert AS_{0}C \rangle$  ($\vert CS_{0}A \rangle$, respectively), which might be defined as a charge separated state. 

The derived equations for the individual matrix elements of the diabatic Hamiltonian are given in figure \ref{fig:Hamiltonian}, expressed in terms of the Fock operator and electron repulsion integrals (ERI) in molecular orbital basis.
The rows' labels correspond to the initial states, whereas the columns' labels represent the final states. In our trimer system, coupling terms emerge involving two-electron integrals that include orbitals from all three monomers, such as  $\langle S_{1}S_{0} S_{0} \vert H \vert S_{0}C A\rangle$ and $\langle C A S_{0} \vert H \vert S_{0}TT \rangle$. We wish to emphasize that these terms do not appear in any dimer model or periodic system which is parametrized by dimer interactions.

The resulting matrix elements can be categorized in those describing the transfer of one electron, which leads to expressions that include both the Fock operator as well as the ERI terms, while the elements involving the transfer of two electrons have only ERI terms. Interestingly, the coupling of $^{1}$TT states with LE and CT states have an negative factor $\sqrt{\frac{3}{2}}$, while the couplings within $^{1}$(TT) states are positive. We then explicitly parametrized the diabatic Hamiltonian for the calculation of diabatic couplings in the considered PDI trimer stack employing quantum chemical procedure described in ("Computational Details"). 

In general, in molecular aggregates with increasing size, the number of electronic pathways grow exponentially. In the trimer system, in total six "direct" and 54 "CT mediated" pathways lead to one of the 2 $^{1}(\text{TT})$ states that have triplet states on adjacent molecules as illustrated in Fig. 2. Following the scheme established by Scholes, in the SF process the transition would then further proceed to the $^{1}(\text{T...T})$ state.
However, also an alternative route can be considered, in which the  $^{1}(\text{T...T})$ intermediate is directly populated from the initial LE state (either in a CT mediated or direct pathway). To the best of our knowledge, such mechanism has not been considered so far.
In order to estimate the efficiency and impact of the individual pathways as well as to determine the structural arrangements that promote the discussed transitions, we carried out 2D scans in the PDI trimer as described in detail in the "Methods" section. 

\subsection{Direct pathway} 
As a first step, we studied the "direct" pathway, evaluating the diabatic matrix elements between the individual LE and the  $^{1}(\text{TT})$ states. The obtained scans are provided in the Supporting Information \cite{supp}. Due to the symmetry of the chosen trimer system, all these scans appear similar, although with opposite signs, exhibiting an extremum located at $\Delta x$=4.0 $\text{\AA}$ and $\Delta y$=0 $\text{\AA}$ with an absolute value of 0.025 eV. However, we wish to point out, that in the expression for the transition probability of the direct pathway, that follows from the first order perturbation theory, the denominator carries the energy difference between the respective diabatic states, leading to an overall negligible SF rate. 
 The matrix elements between the individual LE states and the $^{1}(\text{T...T})$ state are particularly small, of the order of 10$^{-6}$ eV and can be neglected. Notice, that the matrix element $\langle  S_{0} S_{1}S_{0}\vert H \vert TS_{0}T \rangle$ is zero.
 
 \subsection{Charge transfer mediated pathway} 
 We subsequently shifted our attention to the charge transfer (CT) mediated pathway that proceeds from an LE state via a virtual CT state to a $^{1}$(TT) or $^{1}$(T...T) state. As worked out in the Methods part, according to equations 6 and 9, the Transition Probability of the mediated pathway in the trimer stack can be determined by summing over all individual pathways. The result of that scan is presented in \ref{fig:ScanGeometry} a). 
 This plot indicates a highest SF transfer probability at the coordinates $\Delta x= 2.9 \text{\AA}$ and $ \Delta y = 0.0  \text{\AA}$ with an absolute value of 0.025. This rate is relatively low compared  to the individual $T_{LE \rightarrow TT}$, which can be explained by the nullifying effect of the coupling terms of opposing sign in a homo trimer with symmetric slipped stacked geometry (see further discussion below). Consequently, a breaking of symmetry could potentially elevate the CT-mediated coupling.
The  obtained result clearly differs from the SF rate scan obtained for the dimer system (see \cite{miller2017modeling,renaud2013mapping} and our work \cite{Singh2021} ) but resembles the results of the Redfield theory rate scan from Ref.\cite{mirjani2014theoretical}.  However, it should be noticed that in the former cited studies, the resulting rates have been derived by evaluating the matrix elements of only one chosen pathway and therefore do not correspond to  the expressions derived from second order perturbation theory. 

In order to shed more light on the mechanistic details, we studied the individual pathways, focusing on their contribution and structural dependence. 
In particular, we considered the transition from a given LE state situated on one monomer proceeding via any CT state, to i) a given $^{1}$(TT) state on adjacent monomers and ii) the separated $^{1}$(T...T) state. We then determined the transfer probability in accordance with equation \ref{eq:mediatedProbability_CT}. 
Applying these categorization resulted in 9 scans. The corresponding plots are shown in figure \ref{fig:plotMediated}. Each scan in one row shares the same initial LE state, and each scan in a column shares the same final $^{1}$(TT) or $^{1}$(T...T) state, respectively.

The investigated pathways can be further classified in those involving transfer of two electrons, in which one of the correlated triplets in the target state is localized on the monomer that is initially locally excited and in three-electron processes, in which the initially excited monomer does not contribute to the formation of the correlated triplet state. 

For the plots illustrating transitions to $^{1}(\text{TT})$ states (two upper rows in Fig. \ref{fig:plotMediated},) we make the observation, that the scan shape depicting  two-electron processes appear remarkably alike, albeit with opposite signs: The coupling manifests an extremum  at the coordinates $\Delta x= 0.7 \text{\AA}$ and $\Delta y = 0.0  \text{\AA}$   with an absolute value of $\pm0.65$ as well as a local extremum of $\pm$ 0.2, located at $\Delta x$=1.8 $\text{\AA}$, $\Delta y$= 1.5$\text{\AA}$. Interestingly, these plots strongly resemble the scan of the charge trnsfer mediated pathway in the PDI dimer system obtained with the Simple Model or ASD, as reported in our previous work. \cite{Singh2021} 
The plots associated with the three-electron processes, specifically $T_{ S_{0}S_{0}S_{1} \rightarrow TT S_{0}}$, $T_{ S_{1}S_{0}S_{0} \rightarrow S_{0} TT}$ again present striking similarities but exhibit opposite signs. The transfer probability attains its extremum at  $\Delta x= 3.4 \text{\AA}$ and $\Delta y = 0.0  \text{\AA}$ , however it worth noting that the absolute value of $\pm0.008$ is approximately two magnitudes smaller than those of the two-electron process. A further local extremum can be found at $\Delta x$=1.2 $\text{\AA}$, $\Delta y$=0.0 $\text{\AA}$($\pm$ 0.005).

We then studied the formation of the specially separated $^{1}$(T...T) state, considering the three possible pathways, in which the  $\vert T S_{0} T \rangle$ state is formed via a charge transfer-mediated process from one of the three LE states. The resulting scans are shown in the figure \ref{fig:plotMediated} g), h) and i).  $T_{S_{0}S_{1}S_{0} \rightarrow T S_{0} T}$ has a negligible value of the order of $10^{-8}$, bearing a maximum at position $\Delta x= 0.0 \text{\AA}$ and $\Delta y = 0.0  \text{\AA}$. The other two pathways in which $\vert T S_{0} T \rangle$ is formed from those initial LE states, in which the edge fragments' electrons are excited, appear similar but with opposite signs due to the symmetry of the investigated system and exhibit considerable higher values. The scans have a broad extremum between $\Delta x$=3 $\text{\AA}$ and $\Delta x$ =4 $\text{\AA}$ with a center at $\Delta x = 3.4 \text{\AA}, $ and spanning from $\Delta y = 0 \text{\AA}$ $\Delta y =1 \text{\AA}$ with an absolute value of  $\pm$ 0.0010 eV$^2$ . Interestingly, a second basin, located at $\Delta x$=1.5 $\text{\AA}$ and $\Delta y$=1 $\text{\AA}$ shows an additional local maximum with the same value of about  $\pm$ 0.0010 eV$^2$. Notice, that these values are about two magnitudes lower then those of the pathways targeting a $^{1}$(TT) state.

\subsubsection {Individual Pathways} 
Diving deeper into the details, we examined the role of individual CT states in the mediated pathways. These scans are discussed in detail in the Supporting Information.\cite{supp} In general, we discovered that the pathways in which the CT states and the $^{1}$(TT) or the $^{1}$(T...T) states, respectively, are not localised on the same molecules exhibit particularly low probabilities, of the order of $10^{ -7}$ for the former and $10^{-9}$ for the latter.

Highest probabilities are found for those pathways, in which the LE, the CT and a final $^{1}$(TT) state involve the same molecules. Those pathways can be further subcategorized into those where the Anion of the CT state is generated at the initially excited monomer and those where the Cation is located at these position.

\begin{figure}[h]
    \centering
    \includegraphics[width=1.0\linewidth]{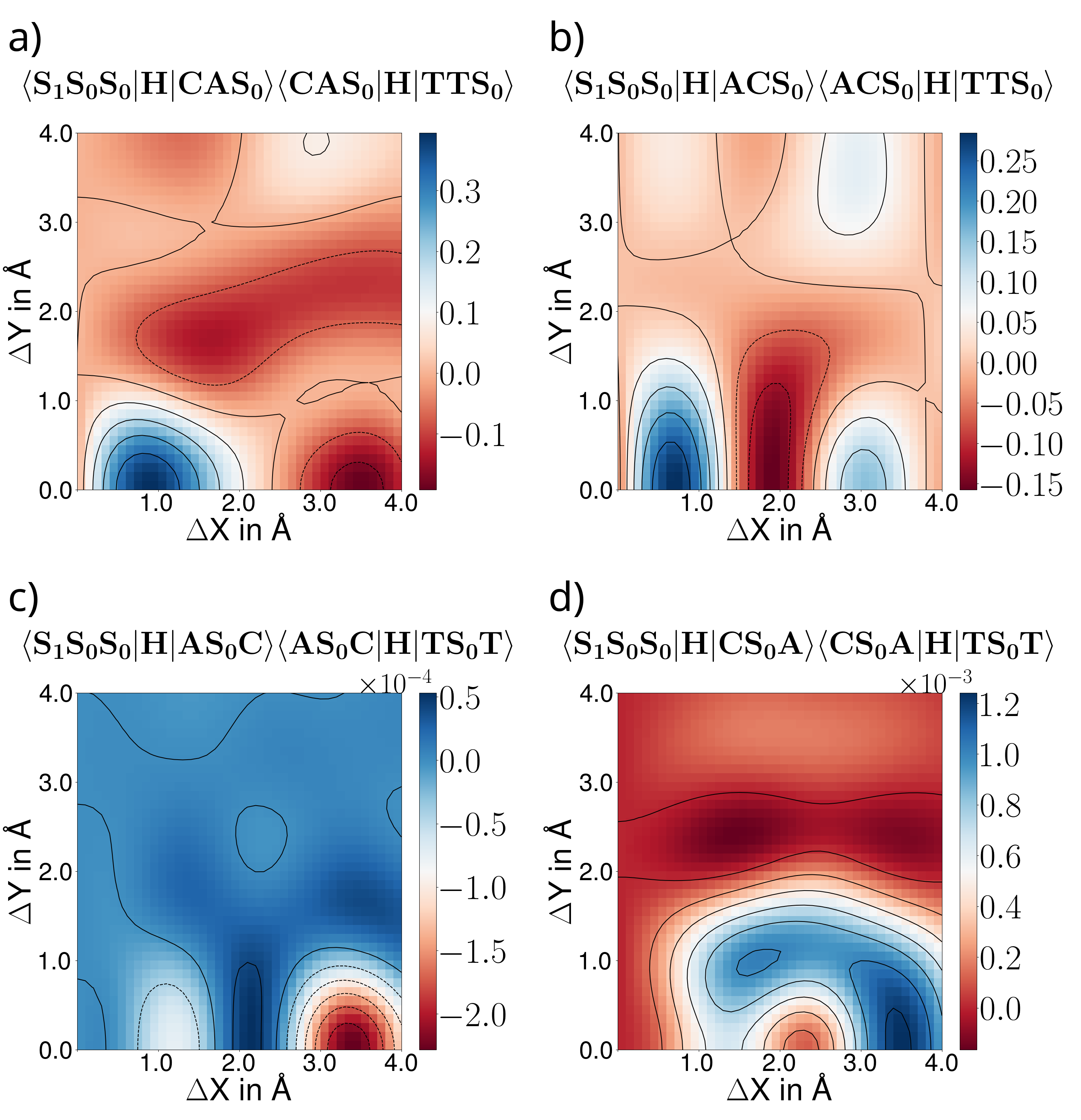}
    \caption{Scans of chosen individual CT-mediated pathways (in eV$^{2}$). For a) and b), CT and $^{1}(\text{TT})$ are located on adjacent monomers. In c) and d),  separated CT and $^{1}(\text{T...T})$ states are considered.  }
    \label{fig:LECTreference}
\end{figure}

For the former, we find a maximum absolute value of about 0.28, while the latter category, the scans exhibit higher values, indicating a maximum at position $\Delta x$ = 0.9 $\text{\AA}$ (abs. value 0.4), as illustrated in \ref{fig:LECTreference}
a) and b). We find, that for these pathways, the initial position of the LE state, whether located at an "edge" molecule or the inner one, does not effect the resulting scans. 
Another category of pathways encompasses those two that involve an LE state on one of the outer molecules, a specially separated C...A state, that is located at the two edge molecules and a specially separated $^{1}$(T...T) state. In this type of pathways, the central monomer can be considered as "innocent" as it remains unexcited throughout the process. 
Again, the resulting scans depend on the position of the Anion (and therefore, also Cation): In case it is localized on the position of the previously locally excited molecule, the absolute probability is smaller, of the order of $10^{-4}$, with an extremum at $3.4\text{\AA}$ (-0.00020). Conversely, for the Cation, the scan reveals a maximum at 0.0012, located at $\Delta x$=3.4$\text{\AA}$, that spans from $\Delta y$=0.0 $\text{\AA}$ to $\Delta y$=1.0 $\text{\AA}$ and a second basin at $\Delta x$=1.8 $\text{\AA}$ and $\Delta y$=1.0 $\text{\AA}$. the overall motif of the scan is closely resembled in  $T_{S_{0}S_{0}S_{1} \rightarrow TS_{0}T}$.  

In the calculation of the individual transfer probabilities, we observe cancellation of matrix elements due to their different signs.  As illustrated in Figure \ref{fig:CancellingMatrix}, the value of the Fock terms of couplings $\langle CAS_{0} \vert H \vert TTS_{0} \rangle$ and $\langle S_{0}AC \vert H \vert S_{0}TT\rangle$ have the same magnitude but opposite signs and so do the Fock terms of $\langle ACS_{0} \vert H \vert TTS_{0}\rangle$ and $\langle S_{0}CA \vert H \vert S_{0}TT\rangle$, while  $\langle CAS_{0} \vert H \vert TTS_{0}\rangle$ and $\langle ACS_{0} \vert H \vert TTS\rangle$ (as well as $\langle S_{0}AC \vert H \vert S_{0}TT \rangle$ and $\langle S_{0}CA \vert H \vert S_{0}TT\rangle$), in which CT and $^{1}$(TT) state are located at the same dimer sub system, have different values. 
In the Transfer Probability, the sum of all products of the individual matrix elements of $\langle$ LE$\vert H \vert $CT $\rangle$and $\langle$CT$\vert H \vert ^{1}$(TT)$\rangle$ are calculated, in total the Fock terms of $\langle$CT$\vert H \vert ^{1}$(TT)$\rangle$ couplings of the two equivalent dimer subsystems have equal value with opposite sign. Altogether the fock terms are canceled. In contrast, the ERI terms of the couplings located on one dimer subsystem, $\langle CAS_{0} \vert H \vert TTS_{0} \rangle$ and $\langle ACS_{0} \vert H \vert TTS_{0} \rangle$ (as well as  $\langle S_{0}AC \vert H \vert S_{0}TT \rangle$ and $\langle S_{0}CA \vert H \vert S_{0}TT\rangle$ on the other dimer subsystem ) exhibit the same value and sign, while equivalent couplings on the "different" dimers differ from each other in sign and values. Fock terms and ERI terms combined give a set of 4 different values of $\langle$CT$\vert H \vert ^{1}$(TT)$\rangle$ couplings shown in figure \ref{fig:CancellingMatrix} c. Summing up all the $\langle LE \vert H \vert CT \rangle \langle CT \vert H \vert TT \rangle$ results in a value of 0.0028eV$^{2}$. 

\subsection {Matrix elements}

We then studied the individual matrix elements between any LE and CT configuration.  First we considered those matrix elements involving CT states on adjacent molecules that share a monomer with an "outer" LE state (in that way, one "edge" monomer remains innocent). The scans available in SI can be classified into two distinct motifs: for those involving a matrix element in which the LE state is situated on the same  molecule as the cation of the CT state, we find a the overall highest absolute values, with a maximum of 0.8 eV, located at the origin, as well as two local extrema, located at $\Delta x$=4 $\text{\AA}$ and Y=0$\text{\AA}$ (0.5 eV) and a broad band spanning from $\Delta x$=0 $\text{\AA}$ to $\Delta x$=4 $\text{\AA}$ and $\Delta y$=1.5 $\text{\AA}$ to $\Delta y$=2.5 $\text{\AA}$ with a center at $\Delta x$=2$\text{\AA}$, $\Delta y$=2$\text{\AA}$ (-0.4 eV). For couplings in which the anion and the LE share the same monomer, the scan also exhibits a global extremum at the $\Delta x$=0 $\text{\AA}$, $\Delta y$=0$\text{\AA}$ position (-0.8 eV) as well as a local one at $\Delta x$=2.5 $\text{\AA}$, $\Delta y$=0 $\text{\AA}$ (0.6 eV). Interestingly, for those transitions involving the LE state on the inner molecule and adjacent CT states, the scans exhibit the same motifs when the Anion or the Cation share the same molecule with the LE state, respectively.
In general, the couplings between the individual LE states and the spatially separated $\vert AS_{0}C \rangle$ (or $\vert CS_{0}A \rangle$) state are weaker, of the order of $1\times10^{-3}$ eV. A clear dependence on the position of the anion and cation can be observed: For the type of couplings with the LE on one of the outer molecules, which is also shared with the cation of the CT state, a maximum at  $\Delta x$=1.2,$\Delta y$=0.8, with a value of $0.011 eV$ is observed. Furthermore, a significant region located between $\Delta x$=3.1 $\text{\AA}$ and $\Delta x$=3.9 $\text{\AA}$ extends in the y direction from 0 to 1 \text{\AA} reaching a value of $0.007 eV$. This motif is partly reflected in the scan for the individual CT-mediated pathway of the $^{1}$(T..T) formation (see discussion above). In the case where LE and Anion share one monomer, the values of the matrix elements are of 1 magnitude lower, with the global extremum at $\Delta x$=3.3, $\Delta y$=0.0 (0.0012 eV). The scans of the couplings between the "inner" LE state and a separated C...A exhibit their extremum located at $\Delta x$=0, $\Delta y$=0 (-0.010 eV) and a local maximum spanning from $\Delta x$=1.8 $\text{\AA}$ to $\Delta x$=3.2 $\text{\AA}$ with a value of 0.002 eV.

We then proceeded by examining the individual matrix elements between the individual CA or C...A and $^{1}$(TT) or $^{1}$(T...T) states, respectively. In general, we find that the couplings have significant values only when the Cation and Anion are situated at the same molecules as the two correlated triplets. Studying these class of matrix elements, we observe, that for the adjacent CT and $^{1}$(TT) states, all the scans  exhibit almost similar patterns: A maximum between $\Delta x$=0.9 $\text{\AA}$ to 1.5 $\text{\AA}$, that spans from $\Delta y$= 0 $\text{\AA}$ to $\Delta y$=0.8 $\text{\AA}$ (0.65 eV) and a local minimum between $\Delta x$=2.9 $\text{\AA}$ and $\Delta x$=3.8 $\text{\AA}$ and $\Delta y$= 0 $\text{\AA}$ and $\Delta y$= 0.5$\text{\AA}$. (-0.4 eV). Furthermore, a broad basin corresponding to a local maximum  spanning from $\Delta y$=3 $\text{\AA}$ (0.2 eV$^2$) and 3.6 (0.45 eV $^2$).
For the transition $\langle AS_{0} C \vert H \vert TS_{0}T\rangle$, a broad maximum is found, with the center spread from at $\Delta x$=2.9 $\text{\AA}$ to $\Delta x$= 4.0 $\text{\AA}$ and $\Delta y$=0.0 $\text{\AA}$ to $\Delta y$=1.0 $\text{\AA}$ with a value of 0.17 eV$^{2}$.

Interestingly, our studies indicate that the regions of strongest LE/CT couplings and CT/TT couplings do not overlap.

\subsection{$^{1}$TT to $^{1}$T...T state transition}

Finally, with the intend to elucidate the process of triplet delocalization and the generation of the  $^{1}$(T...T) state, we performed 2D scans of the Matrix elements once between the two  $^{1}$(TT) states and once between the individual $^{1}$(TT) and the $^{1}$(T...T) state. For the latter, the plot is presented in figure \ref{fig:tripletDelocalisation}.
 The coupling between one conjoint triplet pair to  the other conjoint triplet pair turns out to be very low, of the order $10^{-10}eV$. In contrast, we find that the probability for the formation process from the $^{1}$(TT) intermediate to the $^{1}$(T...T) state based on the matrix elements  $\langle TT S_{0} \vert H \vert TS_{0}T \rangle$ and $\langle S_{0} TT \vert H \vert TS_{0}T \rangle$ has significant values of the order $10^{-2} eV$, with the broad maximum (0.0175 eV) centered at the perfectly stacked geometry. This motif has been previously assigned to the excimer structure in PDIs.\cite{fink_exciton_2008}  While the excimer formation is accompanied with a significant structural relaxation leading to significantly shorter stacking distances\cite{hoche_mechanism_2017}, which we do not account for in the two-dimensional plots, by scanning the individual $\langle$ LE$\vert H \vert$CT $\rangle$ couplings, we have found that the corresponding matrix elements exhibit significant values for the particular structural motif. However, the scans of the $\langle$CT$\vert H \vert^{1}$(TT)$\rangle$ and $\langle CS_{0}A  \vert H\vert TS_{0}T \rangle$ matrix elements indicate a particularly weak coupling for the given structural motif, which would as a consequence hinder the process of SF.   Further studying the scans of the matrix elements $\langle S_{0} TT \vert H \vert TS_{0}T \rangle$,  we find a further broad local minimum ranging from  $\Delta x = 2.2 \text{\AA}$ to $\Delta x = 3.0 \text{\AA}$ with a value of $0.0075 eV$. Interestingly, the plot of the absolute value clearly indicates its center located at $\Delta x$ =2.9 $\text{\AA}$, $\Delta y$=0.0 $\text{\AA}$ which has been also identified as the stacking coordinates corresponding to the maximum of the total Transition Probability for the CT mediated pathway of the $^{1}$(TT) state formation (see discussion above). This finding hints for an efficient $^{1}$(TT) formation. Again, for this region of stacking coordinates, we find significant values for the $\langle$LE$\vert H \vert$CT$\rangle$ coupling, in particular for  $\langle S_{1}S_{0}S_{0} \vert H \vert ACS_{0} \rangle$ and  $\langle S_{1}S_{1}S_{0} \vert H \vert CAS_{0} \rangle$, that hint for the presence of a "local excimer". However, for this structure we find a narrow overlap between significant values of the  respective $\langle$LE$\vert H \vert$CT$\rangle$ couplings and the respective $\langle$CT$\vert H \vert^{1}$(TT)$\rangle$ couplings for the structure with the stacking coordiantes $\Delta x$=2.9$\text{\AA}$, $\Delta y$=0.0$\text{\AA}$ which also exhibits significant coupling to $^{1}$(T...T). Notice, that for this coordinates, also a process via a "separated" C...A state shows overlap of the individual matrix elements. However, as discussed above, these couplings are in general of several magnitudes lower.

\section{Conclusion}\label{sec13}
\begin{figure}[h]
    \centering
    \includegraphics[width=1.0\linewidth]{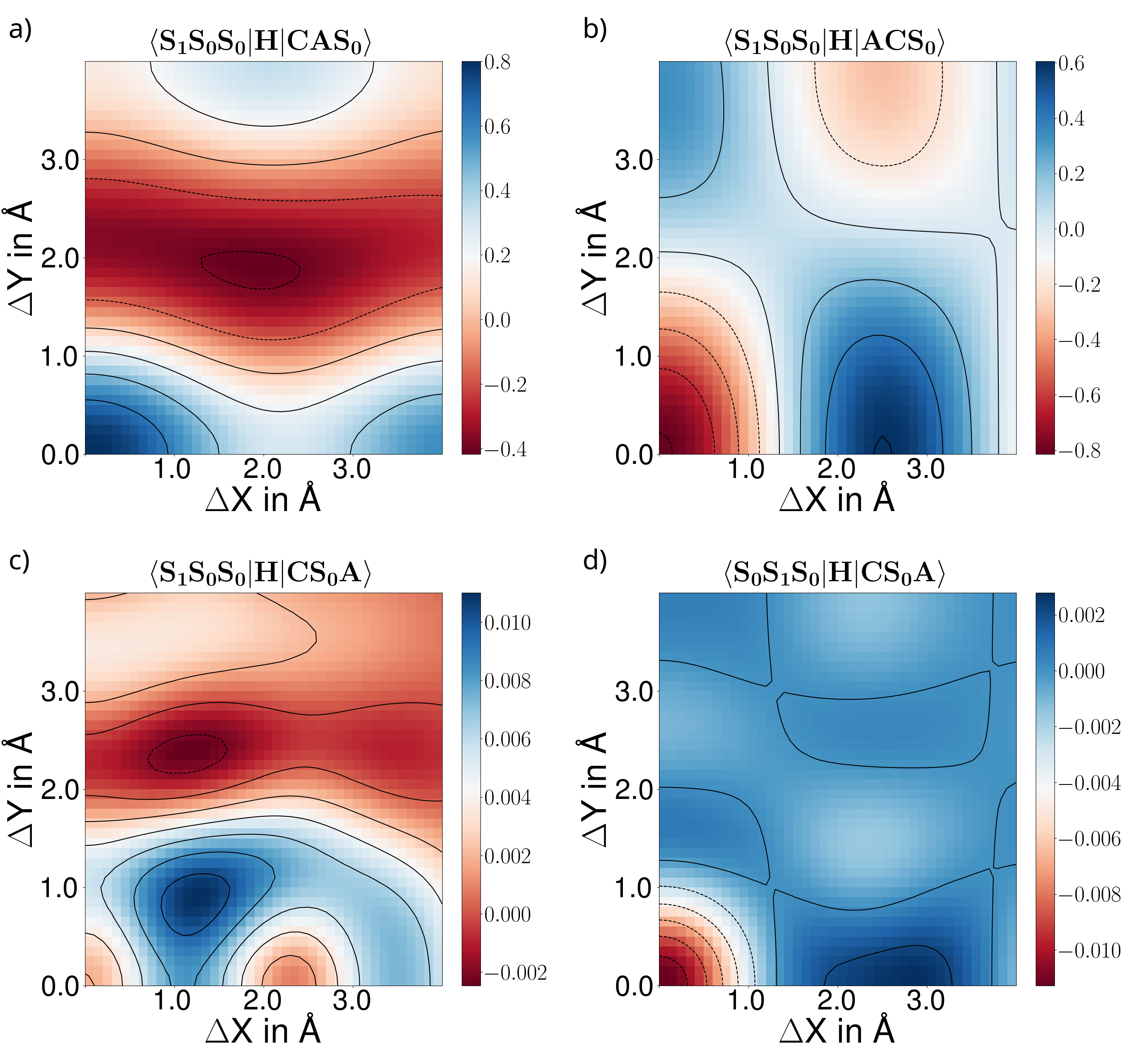}
    \caption{Plot of coupling among LE and CT states (in eV). a) and b) illustrate  matrix elements between an "edge" LE state and CT states located on adjacent monomers.In c) and d) scans of the diabatic coupling between LE to a "separate" CT state.}
    \label{fig:X}
\end{figure}

\begin{figure}[h]
    \centering
    \includegraphics[width=1.0\linewidth]{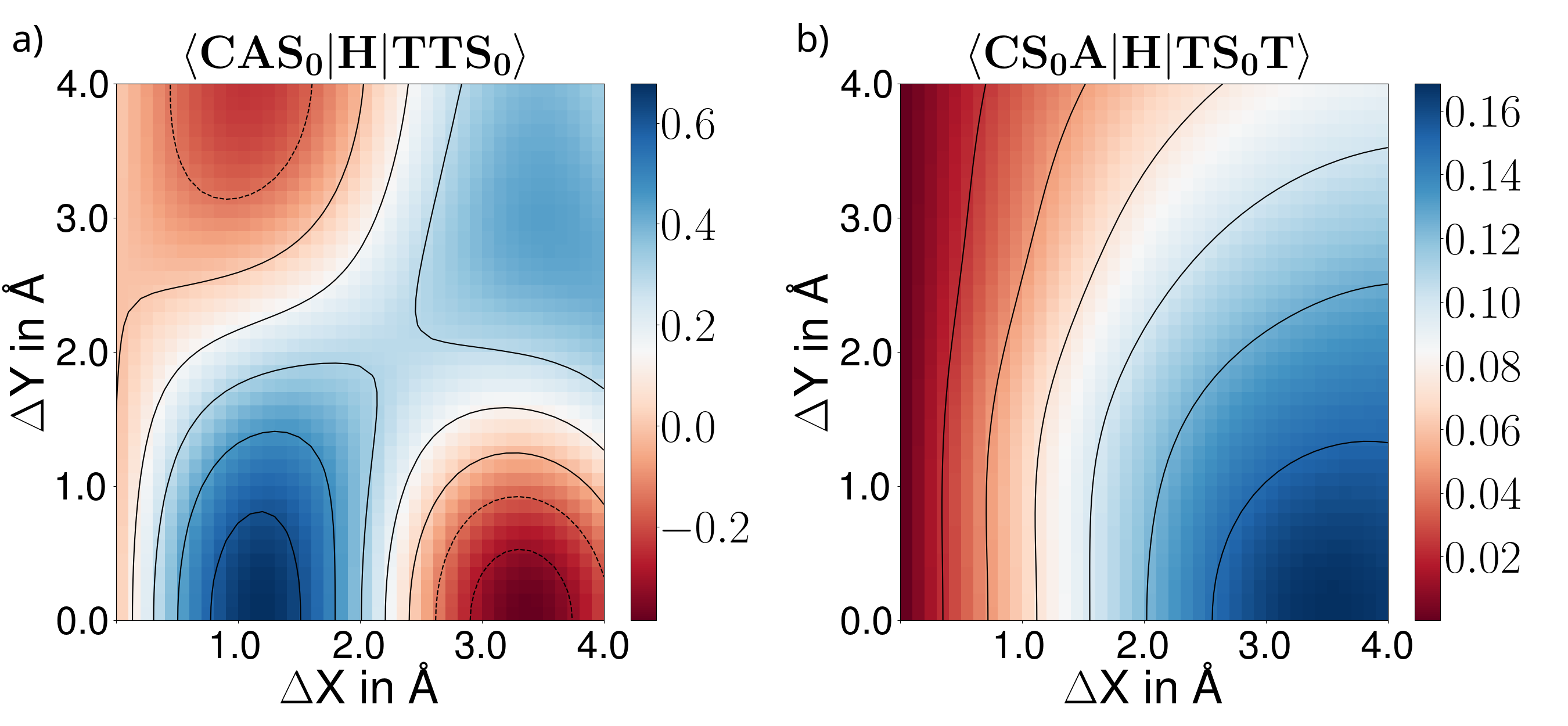}
    \caption{Plot of couplings (in eV) between  "separate" CT state and a) joint triplet state and b) separated triplet state.}
    \label{fig:Y}
\end{figure}

\begin{figure}[h]
    \centering
    \includegraphics[width=1.0\linewidth]{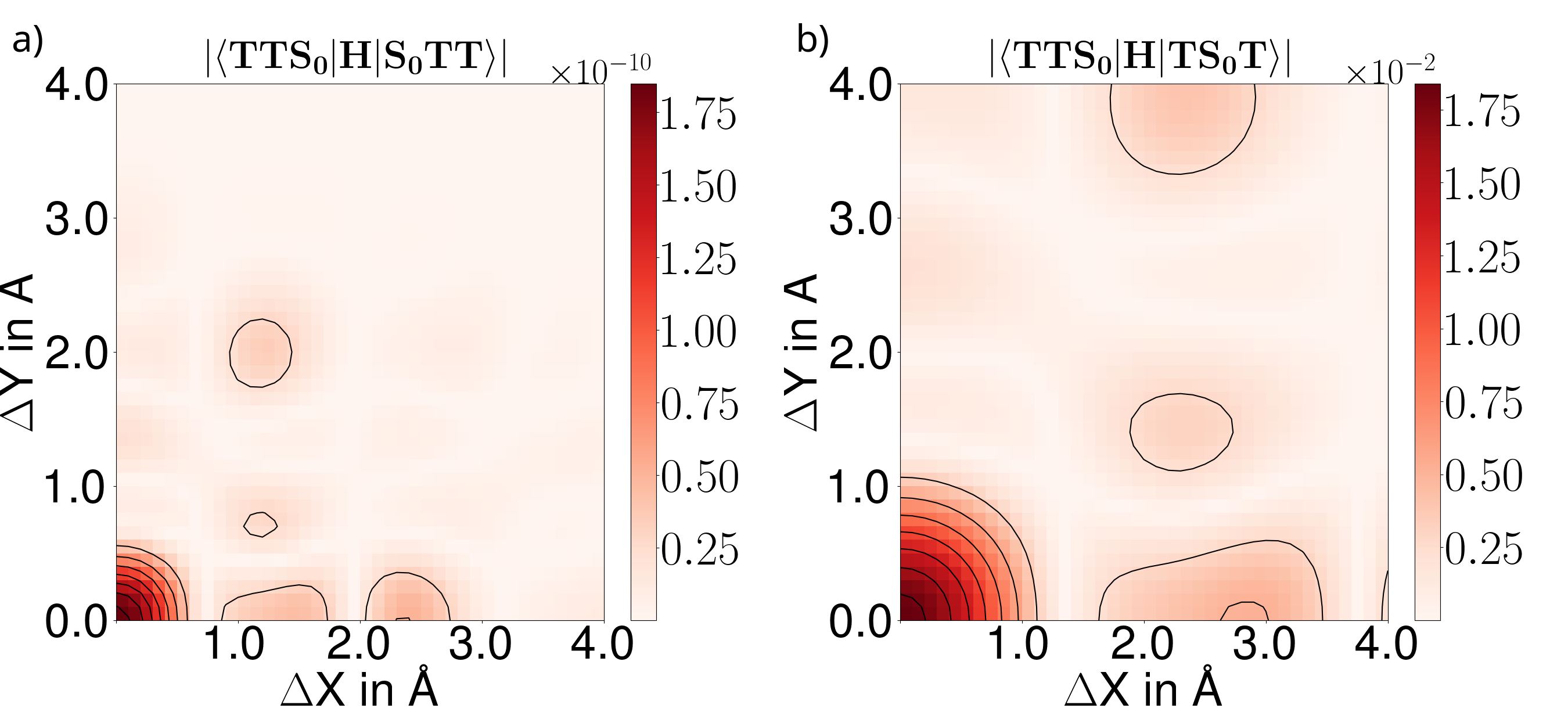}
    \caption{Scans of the absolute value of the coupling between a) the two $^{1}$(TT) states and b) between $^{1}$(TT)and $^{1}$(T...T) (in eV). 
    }
    \label{fig:tripletDelocalisation}
\end{figure}

In this work, we have extended the widely applied diabatic frontier orbital model to the description of trimer systems in order to study the SF process in molecular trimer stacks. Our approach involves explicit parameterization of couplings between all monomers, incorporating also three-center couplings. This allows us to deliver a theoretical study of the formation pathways and their structure dependence of the separated, singlet-spin correlated triplet pair state, $^{1}$(T...T), that is discussed to play a decisive role in the SF process. We focused on the investigation of the trimer systems of cofacially stacked PDIs, that represent a prominent class of SF molecules, and evaluated the impact of the individual electronic pathways and diabatic matrix elements leading to $^{1}$(T...T) formation as well as their dependence on the packing motif by scanning along a longitudinal and a transversal slipping mode. We explored two possible mechanisms: the "two intermediates" scheme, as suggested by Scholes and others, in which a $^{1}$(T...T) is formed upon triplet energy transfer from the singlet spin coupled adjacent triplet pair state $^{1}$(TT) and in addition, a so far unreported mutual pathway that proceeds in a CT mediated process from an initially excited LE state directly to the $^{1}$(T...T). For the latter, we find that the electronic pathway become significant only, when the process proceeds via a "charge separated" state, in which the cation and anion are separated and therefore located on the non-adjacent edge molecules. However, its transition probability is substantially lower compared to the competing $^{1}$(TT) formation pathways. 
 We have then studied the packing dependence of the $\langle^{1}$(TT)$\vert H \vert ^{1}$(T...T)$\rangle$ coupling. Interestingly, the resulting scan shows similarities with the scan of matrix element $\langle S_{0}S_{1}S_{0} \vert H \vert CS_{0}A \rangle$. The coupling is strongest at the perfectly stacked packing motif ( $\Delta x$=0.0 $\text{\AA}$, $\Delta y$=0.0 $\text{\AA}$), another region of significant coupling is found at $\Delta x$=2.9 $\text{\AA}$, $\Delta y$=0.0 $\text{\AA}$, that matches the structural configuration, for which the highest $^{1}$(TT) formation rate has been determined by Redfield theory simulations. Interestingly, we observe that the $\langle ^{1}$(TT)$\vert H \vert^{1}$(T...T)$\rangle$ coupling is significant in regions of strong $\langle$LE$\vert H \vert$CT$\rangle$ coupling corresponding to a "global" and a "local" excimer. 
 
Additionally, employing second order perturbation theory, we screened the SF rate of the $^{1}$(TT) formation in the trimer system, demonstrating that these simple approach reproduces the results of the advanced Redfield theory studies carried out by Miryani et. al.. We find, that the symmetry present in the system leads to cancellation of matrix elements and the total Transition Probability can be mainly broken down to contributions of the electron repulsion integrals, which are often neglected in the implementation of the FO dimer model. 
The simulations show, that the transfer probability of the $^{1}$(TT) formation is highest if the process proceeds via a two-electron process located on a subset of adjacent dimers, in which the LE state and the Cation of the CT state share the same monomer.  
The $\langle$LE$\vert H \vert$CT $\rangle$ coupling is strongest for the perfectly stacked structure which we assign to as "excimer". At this position, the $\langle$ CT$\vert H \vert^{1}$(TT) $\rangle$ coupling is particularly low, therefore the structural motif might be considered as a trap side hindering the SF process. 
A further "local excimer" region can be found for a longitudinally slipped structure between $\Delta x$=2 $\text{\AA}$ and $\Delta x$=3 at $\Delta y$=0 $\text{\AA}$. 
With this study, we hope to motivate researchers from theory, synthesis and spectroscopy to further explore the formation pathways of the $^{1}$(T...T) state as well as a potential role of a separated CT state ($\vert AS_{0}C \rangle$). 

\begin{acknowledgements}
A. S. and M. I. S. R. are thankful for financial support by the Bavarian State Initiative ‘‘Solar Technologies Go Hybrid’’.
\end{acknowledgements}





\bibliography{apssamp}

\end{document}